\documentclass[10pt, onecolumn]{article}

								\usepackage{amsmath}
								\usepackage[dvips]{graphicx}
								\usepackage{amssymb}

\begin{document}


\title{Saddle-Node Bifurcation to Jammed State \\for Quasi-One-Dimensional Counter Chemotactic Flow}
\author{Masashi Fujii, Akinori Awazu, and Hiraku Nishimori\\Department of Mathematical and Life Sciences, Hiroshima University, Hiroshima, Japan}



\begin{abstract}
	The transition of a counter chemotactic particle flow from a free-flow state to a jammed state in a quasi-one-dimensional path is investigated.
	One of the characteristic features of such a flow is that the constituent particles spontaneously form a cluster that blocks the path, called a path-blocking cluster (PBC), and causes a jammed state when the particle density is greater than a threshold value.
	Near the threshold value, the PBC occasionally desolve itself to recover the free flow.
	In other words, the time evolution of the size of the PBC governs the flux of a counter chemotactic flow.
	In this paper,
	on the basis of numerical results of a stochastic cellular automata (SCA) model,
	we introduce a Langevin equation model for the size evolution of the PBC that reproduces the qualitative characteristics of the SCA model.
	The results suggest that the emergence of the jammed state in a quasi-one-dimensional counter flow is caused by a saddle-node bifurcation.
\end{abstract}

\maketitle



	Various types of flows arise in processes or phenomena such as pedestrian or car traffic, granular flow,
	internet communication, chemical reaction networks, and conveyance of proteins
	\cite{helbing, capcarrere, kerner, awazu, chowdhury, dahui, hoogendoorn, burstedde, kuang, ebbinghaus, sugiyama}.
	Considerable advancements have been realized with regard to the investigation and analysis of such flows by using computers to first record data and then apply analytical methods.
	In particular, the transition from a free-flow state to a jammed state
	and the mechanism of the dissolution of a jammed state are important issues in such studies \cite{sugiyama}.
	On the other hand, a study of the movement of ants indicates that ants locally communicate with each other using a class of chemicals called "pheromone" with which they form very long trails to enable effective foraging.
	In a trail of ants, a seriously jammed state such as that found in pedestrian or car traffic is rarely found \cite{kunwar1,nishinari, john1, kunwar2, john2}.
	Motivated by this contrast, we have investigated the traffic of self-driven particles interacting through pheromone
	and we have proposed a stochastic cellular automata (SCA) model \cite{fujii} that is an extended version of the bi-directional ant-traffic model developed in \cite{kunwar2, john1}.
	In this study, we have confirmed that the formation, growth, and decay of the path-blocking cluster (PBC) of particles play intrinsic roles in the time evolution of a traffic flow.
	Below, we briefly explain the outline of our SCA model.
	


	\begin{figure}[p]
		\begin{center}
			\includegraphics[width=75mm]{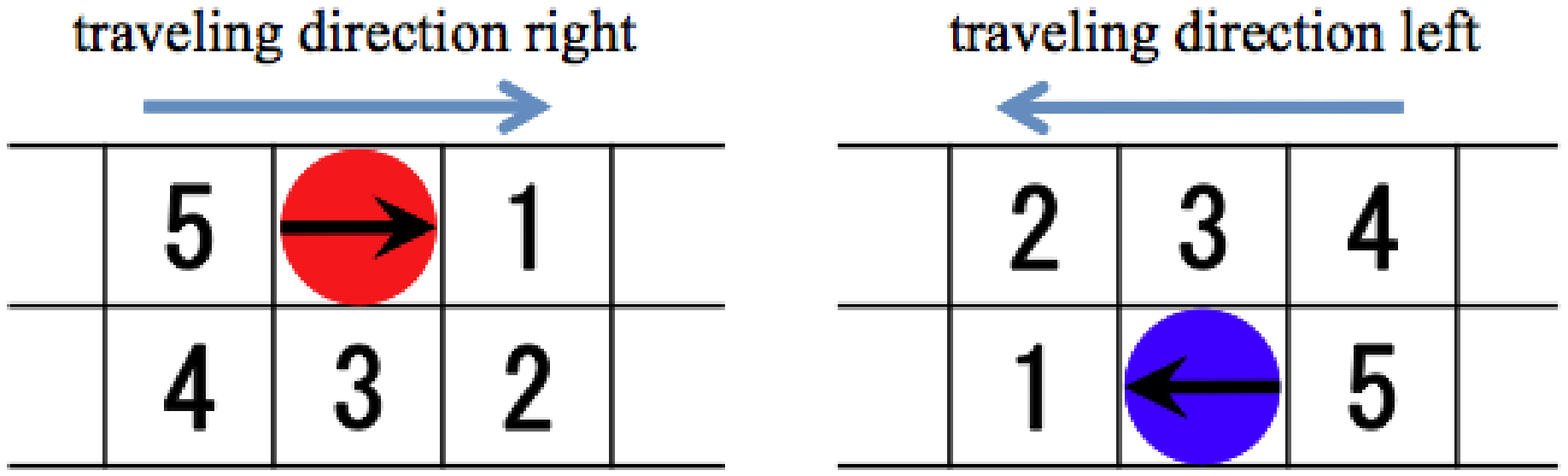}\\[-1pc]
			 \caption{(Color online)
				Ordering for selecting the movement direction of a particle.
			}
			\label{fig:fig1}
			\includegraphics[width=75mm]{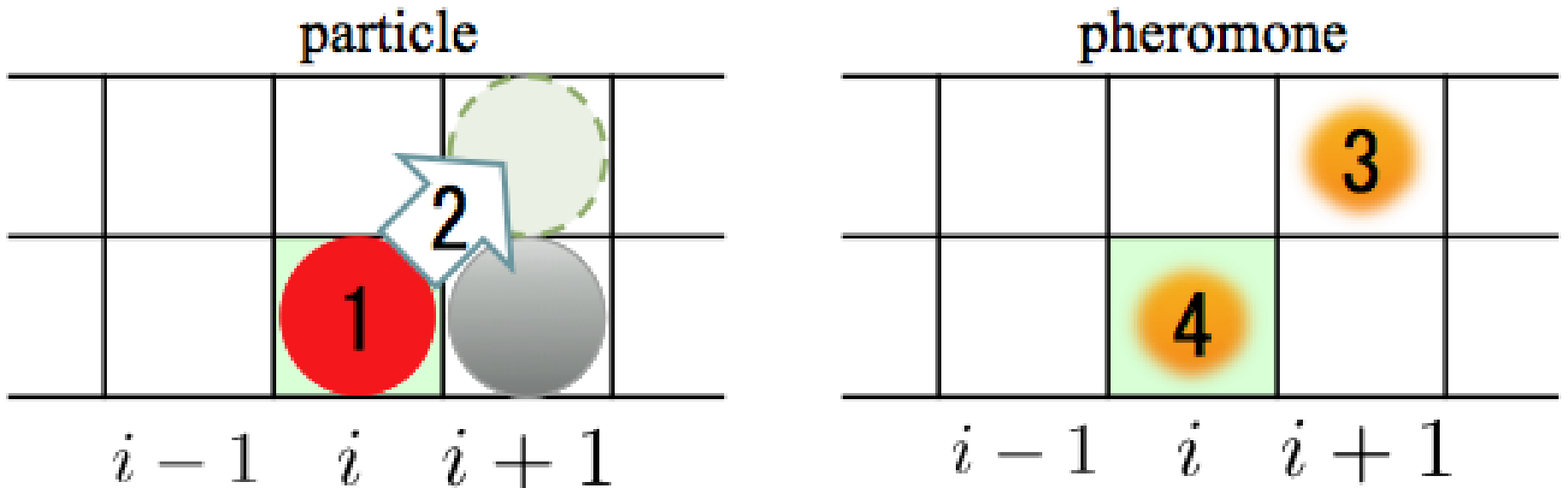}\\[-1pc]
			 \caption{(Color online)
				Updating rules for particles and pheromones.
			}
			\label{fig:fig2}
		\end{center}
	\end{figure}
	The system consists of two lanes, each containing $L$ cells,
	in which particles traveling in both directions share both lanes and swerve to the alternative lane on encountering an oppositely traveling particle in the same lane.
	Each particle has its own traveling direction (right or left).
	When a particle reaches the destination cells at the right or left edge of the system, it reverses its traveling direction.
	In other words, the traveling direction is maintained while the particle travels from one edge to the other.
	Note that the particle movement in an individual step can temporarily be opposite the traveling direction according to the detailed rules explained below.
	To take into account the chemotaxis observed in the traffic of ants \cite{john1, kunwar2},
	particles are set to interact with each other through pheromones; these are stored in a cell for a certain duration.
	In order to introduce the excluded volume effect, only one particle and one unit of pheromone are permitted to be contained per cell at each time step.
	To realize the time evolution of the system, the following updating rules of particles and pheromone fields are introduced (FIG. \ref{fig:fig2});
	\renewcommand{\labelenumi}{\Roman{enumi}) }
	\begin{enumerate}
		\item A cell is randomly selected from among $2\times L$ cells (this random selection is the reason why we call the present CA model "stochastic CA."
		\item If the selected cell is occupied by a particle, a movement direction of the particle is selected according to the rules explained below.
			If a pheromone is left on the neighboring cell in the selected direction, the particle moves to the cell with probability $Q$,
			and in the opposite case, with probability $q$, where $Q>q$ is assumed to reflect the attractive force of pheromone.
			On the other hand, if the cell is unoccupied by a particle, skip step 3 and go directly to step 4.
		\item After the particle moves to a new cell according to step 2, a unit amount of pheromone is inserted into the cell.
		\item The pheromone in the cell from which the particle moves is removed with probability $f$.
		\item Steps 1--4 are repeated $2\times L$ times.
	\end{enumerate}
	All of the above processes constitute a unit Monte Carlo time step.
	The selection rule of the movement direction mentioned in step 2 is given with the following ordering of priority (FIG. \ref{fig:fig1}):
	(1) forward cell,
	(2) diagonally forward cell,
	(3) side cell,
	(4) diagonally backward cell, and
	(5) backward cell.
	If the cell in the selected direction is occupied by another particle, the movement direction is changed to the next priority.

	\begin{figure}[p]
		\begin{center}
			\includegraphics[width=75mm]{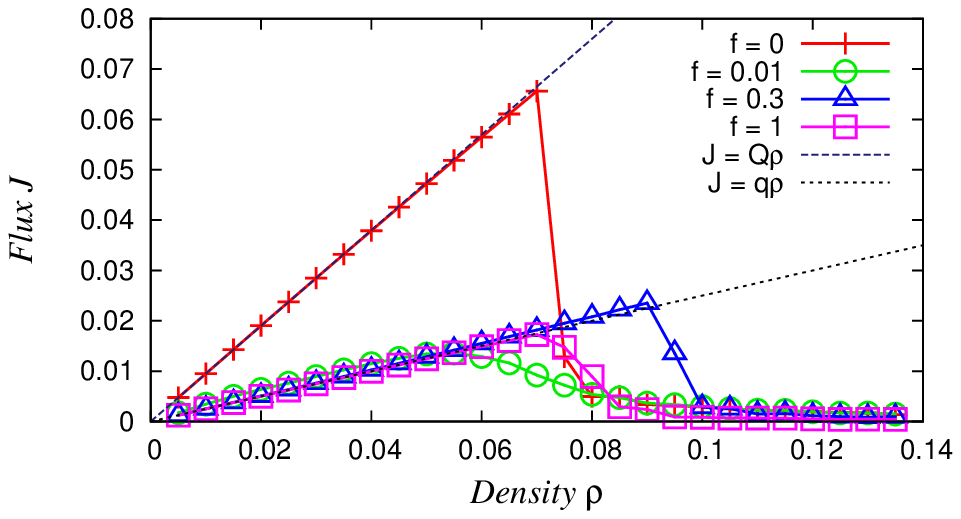}
			 \caption{(Color online)
				$J$-$\rho$ relationship
				for various pheromone evaporation rates: $f$ = 0 ($+$: red), 0.01 ($\bigcirc$: green), 0.3 ($\triangle$: blue), and 1($\square$: purple).
			}
			\label{fig:fig3}
			\includegraphics[width=75mm]{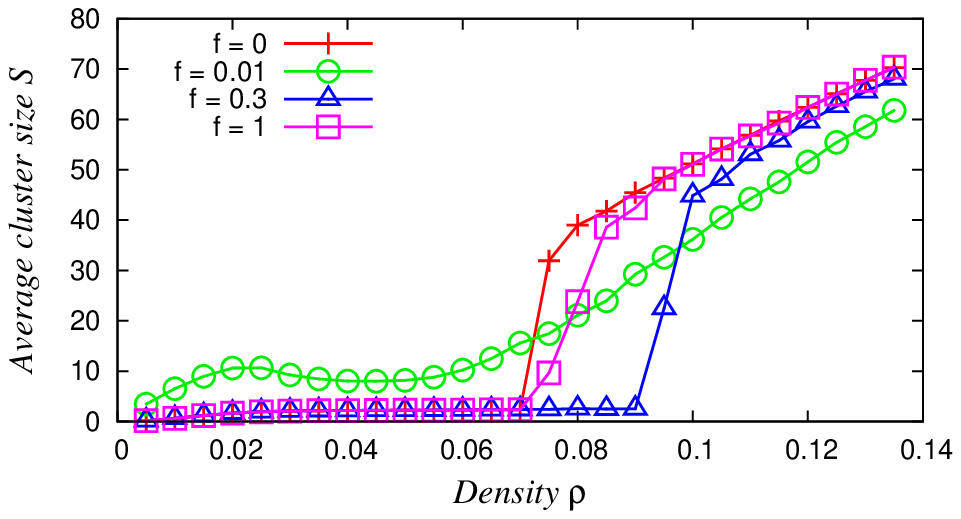}
			 \caption{(Color online)
				$S$-$\rho$ relationship
				for various pheromone evaporation rates: $f$ = 0 ($+$: red), 0.01 ($\bigcirc$: green), 0.3 ($\triangle$: blue), and 1($\square$: purple).
			}
			\label{fig:fig4}
		\end{center}
	\end{figure}
	In our previous study using the SCA model, the flux $J$ and the average cluster size $S$ were numerically investigated. The flux $J$ was defined as the number of times particles traveling leftward reached the left edge (averaged over time steps $0-10^7$) per unit time, and a cluster was defined as a group of more than one particles that is inseparable by more than one contiguous empty column.
	A column was defined as a pair of neighboring cells arranged perpendicular to each lane,
	and the cluster size S was defined as the length of columns per cluster averaged over time steps $0-10^7$.
	As the initial condition, $N$ particles were set randomly in the field.
	The values of the parameters were $Q=0.95$, $q=0.25$, and $L=500$.
	Through numerical simulations,
	we measured the relationship between the particle density $\rho$ ($=N/(2L)$) and the flux $J$, as shown in FIG. \ref{fig:fig3},
	and that between the particle density and the average cluster size $S$, as shown in FIG. \ref{fig:fig4}, for four different evaporation rates of pheromone, $f$ = 0, 0.01, 0.3, and 1.
	The behavior of the particles for $f=0$ was similar to that for $f=1$ because all cells are soon occupied by the pheromone in the latter case;
	therefore, it exhibits dynamics that are simply $Q/q$ times accelerated as compared to the former case.

	\begin{figure}[p]
		\begin{center}
			(a)\hspace{-4.0mm}
			\begin{minipage}[t][][b]{75mm}
			\includegraphics[width=75mm]{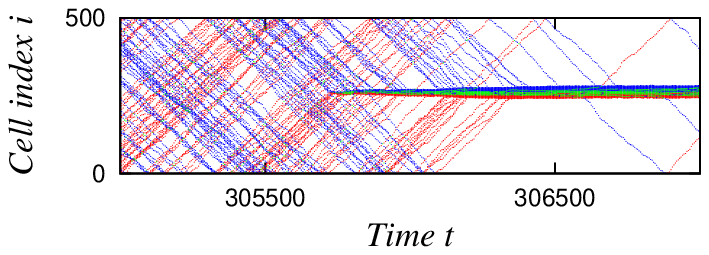}
			\end{minipage}\\
			(b)\hspace{-4.0mm}
			\begin{minipage}[t][][b]{75mm}
			\includegraphics[width=75mm]{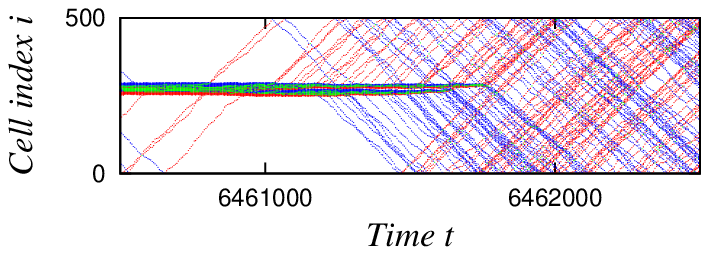}
			\end{minipage}
			 \caption{(Color online)
				Typical spatiotemporal dynamics of particles for $f=0$, $\rho=0.075$, $Q=0.95$, $q=0.25$, and $L=500$:
				(a) formation of a path-blocking cluster (PBC) and (b) dissolution of a PBC.
				Time $t$ (horizontal axis) and column index $i$ (vertical axis) determine the state of a column.
				The red/blue dots indicate that only right-/leftward traveling particle(s) is (are) located in the corresponding columns,
				and the green dots indicate that particles traveling in both directions are contained in the corresponding columns.
			}
			\label{fig:fig5}
		\end{center}
	\end{figure}
	First, as indicated by the $J-\rho$ relationship shown in FIG. \ref{fig:fig3}, for $f$ = 0, 0.3, and 1,
	$J$ linearly increased within a certain density region and decreased below the linear relationship over critical densities depending on $f$.
	Moreover, $J$ at $f$ = 0 was the largest among all evaporation rates of pheromone in the density region $\rho < 0.07$;
	however, $f$ = 0.3 gave the largest $J$ in the density region $\rho > 0.07$, although the average velocity of the particles is close to $q$, because with a certain evaporation rate of pheromone, most of the cells in the low-$\rho$ region do not contain pheromone.
	Next, as observed from the $S-\rho$ relationship shown in FIG. \ref{fig:fig4}, for $f$ = 0, 0.3, and 1,
	the average cluster size $S$ remained almost zero in the low-$\rho$ region in which $J$ linearly increased with $\rho$,
	and $S$ increased sharply when $\rho$ exceeded critical densities depending on $f$.
	On the other hand, for $f$ = 0.01, in the low-$\rho$ region, $S$ was larger than that in other cases; however, it did not have a critical density for the sharp increase, and it continued to moderately increase in the high-$\rho$ region; consequently, $S$ became lesser than that for other values of $f$.
	Furthermore, when $f=0$, 0.3, and 1, in the high-$\rho$ region where $J$ was almost zero and $S$ linearly increased,
	only one cluster including most particles and blocking the path was stably maintained in the system (FIG. \ref{fig:fig5}).
	Hereafter, such a cluster is called a path-blocking cluster (PBC).
	It should be noted that in the intermediate-$\rho$ regions where $J$ decreases for $f=0$, 0.3, and 1, the PBC exhibited finite duration time.
	Namely, the system alternately remained in the jammed and the free-flow states, as clearly observed in FIGs. \ref{fig:fig5} (a) and (b).
	


	In the present study, to realize reduced dynamics of the SCA model by excluding essential features,
	we introduce a Langevin equation for the time evolution of the particle number in the PBC because we consider that the PBC governs the counter chemotactic traffic flow, as discussed below.
	For this purpose, we estimate the leaving and reaching frequencies of particles from and to the PBC, respectively.
	\begin{figure}[p]
		\begin{center}
			\includegraphics[width=75mm]{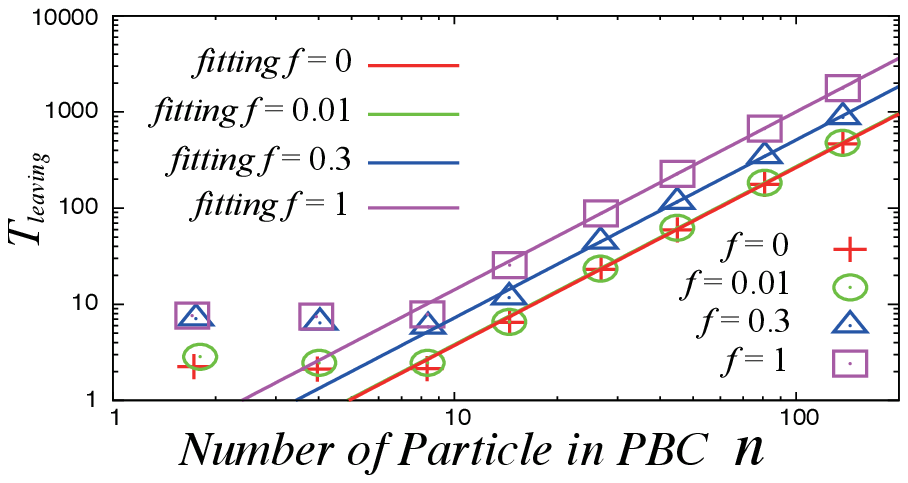}\\[-1pc]
			 \caption{(Color online)
				Relationship between number of particles in the PBC, $n$, and average time interval between successive leaving events of particles from the PBC, $T_{leaving}$,
				for various evaporation rates of pheromone: $f$ = 0 ($+$: red), 0.01 ($\bigcirc$: green), 0.3 ($\triangle$: blue), and 1($\square$: purple).
				The solid lines indicate fitting functions.
			}
			\label{fig:fig6}
		\end{center}
	\end{figure}
	First, to estimate the leaving frequency of particles from the PBC,
	the average time interval between successive leaving events from the PBC (hereafter called the leaving interval, $T_{leaving}$)
	is measured by using the SCA model.
	To maintain the PBC in the simulation of the SCA model, the lane length, $L$, is set to depend on the total number of particles, $N$, such that $L=(N+20)/2$,
	where the moving probabilities $Q$ and $q$ are 0.95 and 0.25, respectively.
	As the initial condition, particles are located at the central region of lanes to form a PBC,
	and the traveling directions of individual particles are set right-/leftward for particles placed at the left/right parts in the central region.
	Through numerical simulation using the SCA model,
	 the relationships between the average number of particles in the PBC $n$ and $T_{leaving}$ are obtained for $f$ = 0, 0.01, 0.3, and 1, as shown in FIG. \ref{fig:fig6}.
	The fitting function for these relationships is given by
	\begin{align}
		T_{leaving}=\frac{1}{20[q+(Q-q)(1-f)^3]}n^{1.85}
		\label{eq:eq1}
	\end{align}
	in the region $n>10$,
	whereas $T_{leaving}$ remains almost constant for $n<10$.
	From this result, the number of particles leaving the PBC per unit time step, $\alpha(n)$, is estimated as a function of $n$ as	
	\begin{align}
		\alpha(n)=
		\left\{
			\begin{array}{ll}	
			20(q+(Q-q)(1-f)^3)n^{-1.85}&(n > 10)\\
			(q+(Q-q)(1-f)^3)/2&(n \le 10).
			\end{array}
		\right.
		\label{eq:eq2}
	\end{align}
	
	Second, the time interval between two successive reaching events to the PBC is estimated (hereafter called the reaching interval, $T_{reaching}$),
	for which purpose the following two assumptions are made:
	(1) only one PBC is formed in the system and
	(2) particles are uniformly distributed outside the PBC.
	When the number of particles in the PBC is $n$ in the present case of two lanes,
	the reaching interval is given by two parameters, the length of lanes $L$ and the average velocity of particles $v$, as
	\begin{align}
		T_{reaching}=\frac{L-\frac{n}{2}}{v}.
		\label{eq:eq3}
	\end{align}
	Consequently, the number of particles reaching the PBC per unit time step is given by
	\begin{align}
		\frac{2v(N-n)}{2L-n}.
		\label{eq:eq4}
	\end{align}
	where $N$ is the total number of particles in the system and the average velocity of particles $v$ is determined as
	\begin{align}
		v=\left\{
		\begin{array}{ll}
			Q=0.95	&{\mbox{if}}\ f = 0\\
			q=0.25	&{\mbox{otherwise}}
		\end{array}\right.
		\label{eq:eq5}
	\end{align}
	because of the behavior of particles obtained in a previous study \cite{fujii}.
	
	From the above discussions, by using the numerical results of the SCA model,
	the time evolution of the number of particles in the PBC is given by
	\begin{align}
	n_{t+1}-n_{t}&=\frac{2v(N-n_t)}{2L-n_t}-\alpha(n_t),
		\label{eq:eq6}
	\end{align}
	\begin{figure}[p]
		\begin{center}
			\includegraphics[width=75mm]{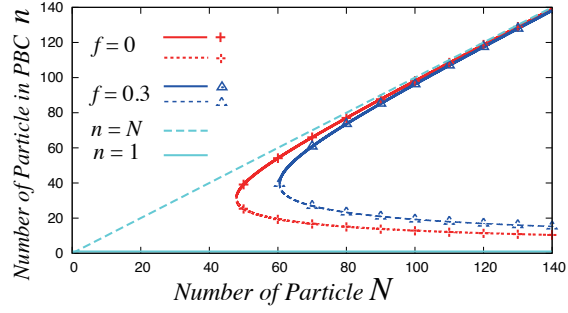}\\[-1pc]
			 \caption{(Color online)
				Relationship between the total number of particles $N$ and fixed points of Eq. (\ref{eq:eq5})
				for $f$ = 0 ($+$: red) and 0.3 ($\triangle$: blue) with $L=500$.
				Solid lines indicate the set of stable points and dashed lines, the set of unstable ones. 
			}
			\label{fig:fig7}
		\end{center}
	\end{figure}
	where $t$ denotes a discrete time step.
	The relationship between the total number of particles $N$ and fixed points of Eq. (\ref{eq:eq5})
	for $f$ = 0 and 0.3 is shown in FIG. \ref{fig:fig7},
	where solid lines indicate stable fixed points and dashed lines, unstable fixed ones; however,
	the fixed points for $f=1$ are not shown because these overlap those for $f=0$.
	The obtained relationships shown in FIG. \ref{fig:fig7} suggest that in the system described by Eq. (\ref{eq:eq5}),
	the PBC emerges as a result of a saddle-node bifurcation if $N$ exceeds a critical value.
	This qualitative behavior appears to be independent of the value of $f$, i.e., the evaporation rate of pheromone;
	however, the critical number of particles for the saddle-node bifurcation depends on $f$.
	Next, to explictly introduce the stochastic effect considered in the SCA model, a Gaussian white noise term is added to the right-hand side of Eq. (\ref{eq:eq5}),
	and we obtain a Langevin equation model:
	\begin{align}
		\dot {n}(t)=\frac{2v(N-n(t))}{2L-n(t)}-\alpha(n(t))+\xi(t),\nonumber\\
		\langle \xi(t) \rangle =0,\ \langle \xi(t) \xi(s) \rangle = 2\Gamma \delta(t-s),
		\label{eq:eq7}
	\end{align}
	\begin{figure}[p]
		\begin{center}
			\includegraphics[width=75mm]{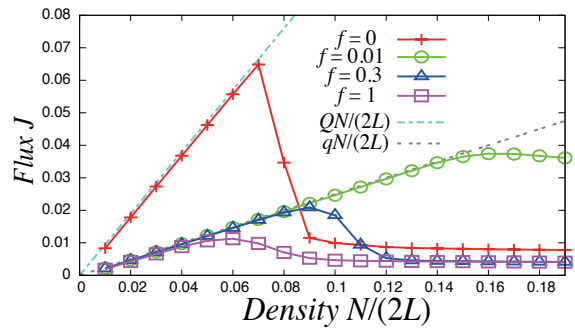}\\[-1pc]
			 \caption{(Color online)
				Relationship between the particle density $\rho (=N/(2L))$ and the flux $v(N-\overline{n}(t))/(2L)$ obtained by the numerical simulation of the Langevin equation model
				for various evaporation rates of pheromone: $f$ = 0 ($+$ :red), 0.01 ($\bigcirc$: green), 0.3 ($\triangle$: blue), and 1($\square$: purple).
				$v$ is set as $Q$ for $f$ = 0, and $q$ for other cases.
			}
			\label{fig:fig8}
		\end{center}
	\end{figure}
	where the bracket $\langle~\rangle$ indicates the ensemble average and $\Gamma$ is a positive number for the index of noise amplitude.
	At the initial time, i.e., $t = 0$, $n(t=0)$ is randomly set in the range from 1 to $N$;
	then, the long-term average of $n(t)$, $\overline{n}(t)$, over time step $10^8$ is obtained by numerical simulation.
	To compare the behavior of the Langevin equation model of Eq. (\ref{eq:eq7}) with those obtained by the SCA (FIG. \ref{fig:fig3}),
	the relationship between the density $N/(2L)$ and the flux $v(N-\overline{n}(t))/(2L)$ is plotted in FIG. \ref{fig:fig8},
	where the values of the parameters are $L=500$, $Q=0.95$, $q=0.25$, and $\Gamma=0.245$.
	It is found that the $J-\rho$ relationships in FIG. \ref{fig:fig3} and FIG. \ref{fig:fig8} qualitatively correspond to each other for $f=0, 0.3$, and 1, although the exact values differ.
	On the other hand, for $f$ = 0.01, the obtained flux continues to increase to a much larger density than that obtained in the SCA model.
	This difference in behavior is due to the failure of assumptions (I) and (II), because the particles easily form more than one PBC for $f$ = 0.01.
	In addition, the flux does not decrease to zero and remains finite for all $f$ in the high-density region.
	This phenomena is caused by the large fluctuation in the high-$n(t) (<N)$ region.



	To summarize, we showed that the characteristic behaviors of a quasi-one-dimensional counter chemotactic flow are governed by the size evolution of a cluster that blocks a path, called a path-blocking cluster (PBC),
	and that this is described by a Langevin equation model.
	This Langevin equation model gives a fundamental diagram similar to that obtained in a previous study of stochastic cellular automata model and it suggests
	 that the transition from the free-flow state to the jammed state in a counter traffic flow can be interpreted as a saddle-node bifurcation on the size of the PBC.
	


	The authors are grateful to members of the Mathematical Society of Traffic Flow for the useful discussions and the information that they provided.
	This study was supported in part by The Global COE Program G14 (Formation and Development of Mathematical Sciences Based on Modeling and Analysis) of the Ministry of Education, Culture, Sports, Science and Technology of Japan.



\end{document}